\documentclass[9pt,twocolumn,twoside]{revtex4-2}

\usepackage{graphicx}
\usepackage[T1]{fontenc}
\usepackage{amsmath} 

\newcommand{\dd}{\mathrm{d}}

\begin{document}

\title{Small field chaos in spin glasses: universal predictions from the ultrametric tree and comparison with numerical simulations}

\author{Miguel~Aguilar-Janita}
\email{miguelaj@unex.es}
\affiliation{Complex Systems Group, Universidad Rey Juan Carlos, 28933 M\'ostoles, Madrid, Spain}
\thanks{\textbf{Author contribution} SF and GP designed research; 
 MAJ, SF, VMM, JMG, GP, FRT and JJRL performed research; 
 MAJ, SF, VMM, JMG, GP, FRT and JJRL analyzed data;  
 MAJ, SF, VMM, JMG, GP, FRT and JJRL wrote the paper.}
\author{Silvio~Franz}
\affiliation{Universit\'e Paris-Saclay, CNRS, LPTMS, 91405, Orsay, France}
\thanks{\textbf{Author contribution} SF and GP designed research; 
 MAJ, SF, VMM, JMG, GP, FRT and JJRL performed research; 
 MAJ, SF, VMM, JMG, GP, FRT and JJRL analyzed data;  
 MAJ, SF, VMM, JMG, GP, FRT and JJRL wrote the paper.}
\author{Victor~Martin-Mayor}
\affiliation{Departamento de F\'{\i}sica Te\'orica, Universidad Complutense, 28040 Madrid, Spain}
\thanks{\textbf{Author contribution} SF and GP designed research; 
 MAJ, SF, VMM, JMG, GP, FRT and JJRL performed research; 
 MAJ, SF, VMM, JMG, GP, FRT and JJRL analyzed data;  
 MAJ, SF, VMM, JMG, GP, FRT and JJRL wrote the paper.}
\author{Javier Moreno-Gordo}
\affiliation{Instituto de Biocomputaci\'on y F\'{\i}sica de Sistemas Complejos (BIFI), 50018 Zaragoza, Spain}
\affiliation{Departamento de Fìsica Te\'orica, Universidad de Zaragoza, 50009 Zaragoza, Spain}
\affiliation{Departamento de F\'{\i}sica, Universidad de Extremadura, 06006 Badajoz, Spain}
\affiliation{Instituto de Computaci\'on Cient\'{\i}fica Avanzada (ICCAEx), Universidad de Extremadura, 06006 Badajoz, Spain}
\thanks{\textbf{Author contribution} SF and GP designed research; 
 MAJ, SF, VMM, JMG, GP, FRT and JJRL performed research; 
 MAJ, SF, VMM, JMG, GP, FRT and JJRL analyzed data;  
 MAJ, SF, VMM, JMG, GP, FRT and JJRL wrote the paper.}
\author{Giorgio~Parisi}
\affiliation{Dipartimento di Fisica, Sapienza Universit\`a di Roma,  CNR-Nanotec, Rome Unit, and INFN, Sezione di Roma1, 00185 Rome, Italy}
\thanks{\textbf{Author contribution} SF and GP designed research; 
 MAJ, SF, VMM, JMG, GP, FRT and JJRL performed research; 
 MAJ, SF, VMM, JMG, GP, FRT and JJRL analyzed data;  
 MAJ, SF, VMM, JMG, GP, FRT and JJRL wrote the paper.}
\author{Federico~Ricci-Tersenghi}
\affiliation{Dipartimento di Fisica, Sapienza Universit\`a di Roma,  CNR-Nanotec, Rome Unit, and INFN, Sezione di Roma1, 00185 Rome, Italy}
\thanks{\textbf{Author contribution} SF and GP designed research; MAJ, SF, VMM, JMG, GP, FRT and JJRL performed research; MAJ, SF, VMM, JMG, GP, FRT and JJRL analyzed data; MAJ, SF, VMM, JMG, GP, FRT and JJRL wrote the paper.}
\author{Juan J.~Ruiz-Lorenzo}
\affiliation{Departamento de F\'{\i}sica, Universidad de Extremadura, 06006 Badajoz, Spain}
\affiliation{Instituto de Computaci\'on Cient\'{\i}fica Avanzada (ICCAEx), Universidad de Extremadura, 06006 Badajoz, Spain}
\thanks{\textbf{Author contribution} SF and GP designed research; MAJ, SF, VMM, JMG, GP, FRT and JJRL performed research; MAJ, SF, VMM, JMG, GP, FRT and JJRL analyzed data; MAJ, SF, VMM, JMG, GP, FRT and JJRL wrote the paper.}

\date{\today}

\begin{abstract}
Replica Symmetry Breaking (RSB) for spin glasses predicts that the equilibrium configuration at two different magnetic fields are maximally decorrelated. We show that this theory presents quantitative predictions for this chaotic behavior under the application of a {\sl vanishing} external magnetic field, in the crossover region where the field intensity scales proportionally to $1/\sqrt{N}$, being $N$ the system size. We show that RSB theory provides universal predictions for chaotic behavior: they depend only on the zero-field overlap probability function $P(q)$  and are independent of other system features.  In the infinite volume limit, each spin-glass sample is characterized by an infinite number of states that have a tree-like structure. We generate the corresponding probability distribution through efficient sampling using a representation based on the  Bolthausen-Sznitman coalescent. Using solely $P(q)$ as input we can analytically compute the statistics of the states in the region of vanishing magnetic field. In this way, we can compute the overlap probability distribution in the presence of a small vanishing field and the increase of chaoticity when increasing the field. To test our computations, we have simulated the Bethe lattice spin glass and the 4D Edwards-Anderson model, finding in both cases excellent agreement with the universal predictions.
\end{abstract}


\maketitle

In low-temperature spin glasses, Replica Symmetry Breaking (RSB) theory predicts that for a given large sample, equilibrium states are non-self-averaging and organized as the leaves of a weighted random tree \cite{mezard:84}. The statistics of these trees are universal, they depend only on the overlap order parameter function $q(x)$, or equivalently, the average overlap probability distribution $P(q)$. The relative weights of the branches, proportional to the exponential of the free energies of the states, also display universal statistics, obeying the so-called Ruelle probability cascades \cite{ruelle:87,mezard:85b,derrida1985generalization,panchenko2013ruelle}.  

Despite having first been derived in the Sherrington-Kirkpatrick (SK) model this description
is completely general: it is valid in more general Mean Field models such as spin glasses on finite coordination random graphs\footnote{The SK model has a {\sl simple} RSB solution because the free energy can be computed using a Gaussian stochastic process on the tree. This is not true in finite coordination systems, where the process is no more Gaussian and the RSB solution is more complex.}. If  RSB is correct  for finite-dimensional spin-glass systems, these predictions should be also valid in this case: a failure of these predictions would falsify RSB in its present form.

This complex statistical structure has striking consequences on the behavior of the overlap distribution function and the magnetization as a function of the external magnetic field $H$ in the crossover region where the magnetic field scales with the system size $N$ as $N^{-1/2}$ (the choice of the exponent $1/2$ will be justified later). We find that these quantities can be exactly computed as a function of $h=H N^{1/2}$ in the infinite volume limit: at a fixed value of $h$ the magnetic field $H=h/N^{1/2}$ goes to zero, consequently we call $h$ the vanishing magnetic field. 

We stress that all the analytic predictions are strictly valid in the infinite volume limit; finite volume systems will present corrections that strongly depend on the structure of the system and that are not well understood from first principles. It is clear however that these corrections should be smaller and smaller for large volumes. 

\subsection*{Vanishing fields}
Vanishing external fields are interesting when a system has more than one equilibrium state, adding an external perturbation $\delta {\cal H}$ to the Hamiltonian has two effects:
\begin{enumerate}
    \item Change the relative weights of the equilibrium states; the interesting situation is when these effects are of $O(1)$.
    \item Change the properties inside a given state.
\end{enumerate}
In the ideal scenario, the external perturbation $\delta H$ is strong enough to produce the effect (1), but weak enough to not produce the effect (2). Indeed we can compute (1) from first principles, provided that we know the phase structure of the system and the value of the order parameter, while the computation of (2) requires a much deeper knowledge of the system that often is not 
available. Therefore the scaling with $N$ of the perturbation should be carefully chosen to satisfy both criteria and to produce not-too-simple predictions.

Let us consider a crystal clear case: the Ising ferromagnetic model with symmetric boundary conditions (e.g., periodic boundary conditions). In this model, at low temperatures and zero magnetic field, the equilibrium Gibbs measure is the mixture of two pure states\footnote{A state is pure if  the connected correlations vanish at infinity; for translation invariant systems there are many equivalent definitions of pure states.} with positive and negative magnetization $\pm m$ with equal weights $w_\pm=1/2$.  We can write, 
for an observable $\Omega$ depending on the configuration of the system:
\begin{equation}
    \langle \Omega\rangle_G=w_+\langle \Omega \rangle_+ +w_-\langle \Omega\rangle_-
 \,,
\end{equation}

where the two states $\pm$ are clustering (i.e., the connected correlation functions go to zero at large distances). The pure states can be obtained by adding an external field $H$: we have to consider first the limit where $N$ (the total number of spins) goes to infinity and later the limit  where $H$ goes to zero. Indeed in the limit $H\to 0^\pm$ we obtain the state $\langle \rangle_\pm$ .

All that is well known. We now introduce the vanishing field  $\tilde h\equiv N H$ and we study what happens in the limit of $N\to \infty$ at fixed $\tilde h$
(we write $\tilde h$, rather than $h$, to stress the fact that these arguments refer to a ferromagnetic system; we choose $\tilde h>0$ to fix the ideas). 
With such scaling the perturbation of the Hamiltonian is finite in typical configurations. The field is not strong enough to suppress the '-' states, but it is strong enough to modify the weights, which are now 
given by $w_\pm=(1\pm \tanh(\beta \tilde{h} m ))/2$, where $m$ is the spontaneous magnetization per spin.  We have a crossover behavior as a function of $\tilde{h}$,  and the usual situation with only a single pure state is obtained only for $|\tilde{h}| \gg 1$.

This can be shown using perturbation theory in $\tilde{h}$, that implies that for an observable $\Omega$, 
\begin{equation}
    \langle\exp(-\beta \delta {\cal H}) \Omega\rangle_\pm=
    \exp(\pm \beta \tilde{h}
    m + \frac{\beta^2 \tilde{h}
    ^2}{2}\chi+{\cal O}(\tilde{h}^3/N^{1/2})) \langle \Omega\rangle_\pm\;,
\end{equation}
$\chi$ being the magnetic susceptibility.
The effects of the vanishing field are dominated by the leading term proportional to the spontaneous magnetization $m$; the term $\tilde{h}^2\chi$ is the same in both states and hence it does not change their weights.

With an eye on  the spin glass case that will be considered in the next section, it may be interesting to consider as well for a ferromagnetic system the overlap between the equilibrium configurations of two replicas $\sigma$ and $\tau$
\begin{equation}
    q=\frac{1}{N}\sum_i\sigma_i\tau_i
\end{equation}
The probability distribution of $q$ can be easily computed as function of $\tilde{h}$ 
\begin{eqnarray}
\label{eq:A}
    P(q;\tilde{h})=A_+(\tilde{h})\delta(q-m^2)+A_-(\tilde{h})\delta(q+m^2)\;,\\
    A_\pm(\tilde{h})=\frac12 \left( 1\pm\tanh^2(\beta \tilde{h} m )\right)\;.
\end{eqnarray}
When the vanishing field $\tilde{h}$ becomes large we find $A_+(\tilde{h})\to 1$  and $A_-(\tilde{h})\to 0$.
We can also consider the probability distribution $P_C(q;\tilde{h})$ of the 'chaos overlap' i.e. the overlap between one configuration at $\tilde{h}=0$ and another at $\tilde{h}\ne 0$. 
In this case, independently of $\tilde{h}$, one simply has
\begin{equation}
    P_C(q;\tilde{h})=\frac12 \left( \delta(q-m^2)+\delta(q+m^2) \right)
\end{equation}

We could also consider the effect of a random magnetic field in a ferromagnetic system: $\delta {\cal H}=-\hat{h}/N^{1/2}\sum_{i=1}^N r_i\sigma_i$ where the $r_i$ are centered random variables with variance one. 
If we define the total magnetic field by\
\begin{equation}
    N H=\frac{\hat{h}}{N^{1/2}}\sum_{i=1}^N r_i \,,
\end{equation}  
then the vanishing field for this random case would be
$H=(z \hat{h}/N)$,
where $z$ is a Gaussian number of unit variance.  In this case we have a different power of $N$ in the definition of the vanishing field because, 
in order to have a finite effect on the weights, we need a finite value of $\langle \delta {\cal H}\rangle_\pm $ in the infinite volume limit. 

The effect of such a perturbation is slightly more subtle 
that the one of a constant field. In the presence of the field, the weights of the states are random variables that depend on the $r_i$. One needs now to define $P(q)$ averaging over the $r_i$ $P(q)=\overline{P_r(q)}$. 
A straightforward computation reveals that 
in this case $P(q)$ is still as in (\ref{eq:A}) but with
\begin{equation}
    A_\pm(\hat{h})=\frac12\pm\frac12\int d\mu(z) \tanh^2(\beta h z m )\,,
\end{equation}
where $d\mu(z)$ is a Gaussian probability distribution with unit variance and zero average. 

While these results are simple and perhaps not surprising, 
we fear that no one has spelled them in detail. 
They are valid for any dimensions 
whenever in the low-temperature phase we have a spontaneous magnetization. 
It should be emphasized that this is true even in two dimensions, where 
any macroscopic random field destroys the ordered phase.
In this marginal case perturbation theory breaks down as soon as $\hat{h}=O(N^{1/2})$ and 
our formulae in the vanishing field regime, give no information on the behavior 
when the 
field is $H=O(1)$.
 
\subsection*{Spin glasses}
The situation is more complex in spin-glasses in the standard RSB picture where the number of equilibrium states is infinite. When a small random perturbation is added, the Boltzmann weights of the tree's branches are reshuffled.
As the intensity of the perturbation grows, states that originally had a small weight become dominant, and the overlap
between the unperturbed and the perturbed system
decreases. For any small but macroscopic perturbation, this implies vanishing similarity, a property known as chaotic dependence on the intensity of the 
perturbation~\cite{young:98,ney-nifle:97,kondor:89}.
 
A well-known example of chaotic behavior is chaos against 
temperature~\cite{mckay:82,bray:87b,billoire:00,janus:21}
where, however, a variation of temperature does not result in a random perturbation unrelated to the energy landscape (see later). Quantitative predictions are more difficult and less universal: both kind of spin glass models ---with chaos and without chaos--- are known. Even in mean field models, like the Sherrington-Kirkpatrick model, where chaos is present, explicit computations have only been done near the critical temperature\footnote{Temperature chaos, when present, is due to the subtle energetic-entropic balance in the states, that is affected by temperature changes~\cite{franz1995chaos,rizzo2003chaos,fernandez2013temperature,fernandez2016temperature,chen2017temperature,panchenko2016chaos}.}.

In this paper, we study 
the effect of 
a magnetic field, where the theory is simpler and chaos has universal features that are absent for temperature chaos\footnote
{The numerical study of chaos in a magnetic field was started twenty years ago in \cite{billoire2003magnetic} for the SK model.}. 
 The development of a general theory  
 is possible here
 because 
in the absence of a field, the average magnetizations of the states are Gaussian random variables uncorrelated with the energy of the states. This generic property always holds if there are many states and random couplings without a ferromagnetic component. 

 The interesting zero temperature limit of the present problem, where ground-state level crossings lead to avalanche-like discontinuous jumps in the magnetization, was 
studied for long range models in Refs. \cite{ledoussal:10,ledoussal:12,franz2017mean}.  Using an adaptation of the replica method introduced in Ref. \cite{bouchaud1995scaling} a closed formula for the avalanche density in the limit $h\to 0$ was found. Here 
we are interested in finite temperatures, and also in the full dependence on the spin-glass vanishing field $h=HN^{1/2}$.\footnote{This scaling is easy to understand at small $h$: due to the randomness of the $J$, the model with constant $H$ is equivalent to the model with a random bimodal $H_i$. In this situation in the replica approach, the magnetic field corresponds to a new term in the Hamiltonian, that is equal to
$\frac12 H^2\sum_{i=1,N}\sum_{a,b=1,n} \sigma^a_i \sigma^b_i+O(H^4)$.
The coefficient of the term proportional to $ H^2$ is of ${\cal O}(N)$ so this term is of ${\cal O}(H^2 N=h^2)$.}
 
Remarkably, the predictions of standard RSB about magnetic field chaos can be submitted to a {\it quantitative} test, once the function $P(q)$, or equivalently $q(x)$, at zero magnetic field has been measured. Indeed, one can generate random trees and their relative weights in the absence or the presence of a field, and directly compare with the results of numerical simulations. 

 In this paper, we compare the theory's predictions with the simulations of two models exhibiting low-temperature spin glass behavior:
 the Bethe lattice spin glass and the 4D Edwards-Anderson model. 
Deviations from the theory can be expected either for finite-size effects or because of the absence of standard RSB. To calibrate the first effect, we study the Bethe lattice spin glass, where the standard RSB structure of the states is present (although 
only approximate RSB solutions are available); 
in this case, we find a very good agreement with the theory as expected. In a second moment, we study the 4D spin glass, where qualitatively we find very similar finite size effects. Also, in this case,
the agreement is excellent, 
consistently with clear evidence in favor of RSB in this system at least at zero magnetic field \cite{parisi1996equilibrium,banos2012thermodynamic,charbonneau2023spin}. We are not discussing in this note the existence of RSB in the presence of a fixed non-zero magnetic field $H={\cal O}(1)$. Different viewpoints exist on this separate problem,  (e.g., Ref.~\cite{moore:21}) which is not relevant to us because we are concerned with fields  $H\sim 1/N^{1/2}$.

\subsection*{Broken Replica Symmetry theory}
For the reader's convenience, and to establish notations, let us recapitulate the theory using slightly imprecise but simplified language. We suppose to deal with a spin glass without a net ferromagnetic component in the couplings and invariant under spin reversal. In the RSB phase in each instance of the system, there is an infinite number of equilibrium states labeled by $\alpha$.  These states have random statistical weights $w_\alpha$, normalized to $\sum_{\alpha=1}^\infty w_\alpha =1$, 
 magnetizations
$m_\alpha=N^{-1/2}\sum_{i=1}^{N} \langle  \sigma_i\rangle_\alpha$,
 and mutual overlaps $q_{\alpha \gamma}=N^{-1}\sum_{i=1}^{N}  \langle  \sigma_i\rangle_\alpha\langle  \sigma_i\rangle_\gamma$,
where $\langle (\cdots) \rangle_\alpha$ denotes the statistical average in the state labeled by $\alpha$. The normalization of the magnetization has been chosen to have $m_\alpha=O(1)$ for large $N$. In fact, in each state spins freeze in random directions and the $m_\alpha$ are Gaussian, zero mean variables, with covariance  $\langle m_\alpha m_\gamma\rangle=q_{\alpha\gamma}$. 
The {\sl positive} overlaps are constrained by ultrametricity, implying that the states are organized as random trees, whose statistics we describe later.

It is convenient to define the probability distribution of the overlap for a given disorder realization (sample) $P_J(q)$ and its  average\footnote{This and the following similar equations should be thought of as holding in distribution sense in the thermodynamic limit.}:
 \begin{equation}
 P_J(q)=\sum_{\alpha,\gamma} w_\alpha w_\gamma \delta (q- q_{\alpha\gamma}) \,,\quad P(q)=\overline{P_J(q)}\,,
\end{equation}
being $\overline{(\cdots)}$ the average over the disorder. In mean-field models, $P(q)$ has a continuous component and a $\delta$ peak:
$P(q)=\tilde P(q) +(1-x_M) \delta (q-q_{\mathrm{EA}})$,  with $x(q)=\int_0^{q}\dd q^\prime\,P(q^\prime)$ and $x(q_\mathrm{EA})=x_M$.  Notice that the function $q(x)$ is defined as the inverse function of $x(q)$.

We can ask now what happens to the overlap   when in a given system one replica is at zero magnetic field and the other at non-zero {\sl small} 
magnetic field $H=h/\sqrt{N}$  (both replicas of the system share the same disorder). In this regime one only adds a finite perturbation ---keep in mind that the total magnetization is of order $\sqrt{N}$ which is the rationale for the choice of  normalization in the definition of $m_\alpha$. In this way, the states keep their identity, while their weights are modified: $w_\gamma \to w_\gamma(h) \propto w_\gamma  \exp(\beta h m_\gamma)$
with $\sum_\gamma w_\gamma(h)=1$. Notice that, as it is well known \cite{mezard:85b}, the distribution of the weights is left invariant 
by such a re-weighting independently of the value of $h$.  We are interested here to the correlations between the original weights and their re-weighted version.
We readily find that the probability distribution 
of the overlap between the two replicas of the system (one in the absence and the other in presence of the field) is given by
 $P_{J,C}(q;h)=\sum_{\alpha,\gamma} w_\alpha w_\gamma(h) \delta (q- q_{\alpha\gamma})\,,$
 or, when both replicas of the system feel the same magnetic field,
 $P_{J}(q;h)=\sum_{\alpha,\gamma} w_\alpha(h) w_\gamma(h) \delta (q- q_{\alpha\gamma})\,.$
 
 An additional grain of salt is due, because the system is invariant under spin reversal at $h=0$. For any state $\alpha$, there is a state $\alpha'$ with opposite magnetization and $w_{\alpha'}=w_{\alpha}$. It is convenient to remove this degeneracy: we use only one label for each pair of states and we define $q_{\alpha\gamma}$ such that $q_{\alpha\gamma}>0~\forall (\alpha,\gamma)$. This is the convention we use in the generation of the spin glass trees. In the end one reconstructs 

  \begin{equation}
  P^C_J(|q|;h)=\sum_{\alpha,\gamma} w_\alpha w_\gamma (h) \delta (q- q_{\alpha,\gamma})\,.
  \end{equation}
 where now $w_\gamma(h) \propto w_\gamma  \cosh( \beta h m_\gamma)$.
Similarly, the probability distribution of the overlap (positive  {\it and} negative) between two replicas in a field $P_J(\pm|q|;h)$ is: 

\begin{equation}
    P_J(\pm|q|;h)= \sum_{\alpha,\gamma} w_\alpha(h) w_\gamma(h) g_\pm (h,m_\alpha,m_\gamma) \delta (|q|-q_{\alpha\gamma})
\end{equation}
with $g_\pm(h,m_\alpha,m_\gamma)= [1\pm \tanh(\beta h  m_\alpha) \tanh(\beta h  m_\gamma)]/{2}$. 
Note that in the limit $h \to\infty$ [$H>0$, fixed, while $N \to \infty$]: $ P_C(|q|;\infty)=\delta(q)\,, P(q;\infty)=\theta(q)P(|q|).$

An additional quantity of interest is the average magnetization induced by $h$; 
using the previous formulae we find that $m(h)=\beta h [1-\int_{-1}^1 \mathrm{d}q\, q P(q;h)]$ (see SI for details).

All these functions  are nontrivial(at variance with the case of ferromagnetic systems).  Remarkably,  they can be computed efficiently, provided that we know the function $P(q)$ at zero magnetic field. 

The simplest approach, which we follow here, is to average over a large number of random trees generated with the correct statistical properties\footnote{It will be shown in a further publication how the same results could be obtained by solving partial differential equations associated with diffusion on random trees.}. The reader may find a detailed explanation of how to generate RSB spin-glass trees of states with the right statistics in \textit{Materials and methods}. Furthermore, all the technical details of the generation of the trees can be found in the SI.

As we have stressed an input to this kind of computations is the order parameter, in the ferromagnetic case, it is $m$, in the RSB approach it is $P(q)$. In the SK model, $P(q)$ can be exactly computed analytically. However, this is a lucky exception and in general we have to obtain $P(q)$ from numerical simulations.
As explained above, the probability distributions $P(q)$, $P(q;h)$, and $P_C(q;h)$ will be the main quantities to study. To compare the theoretical predictions with the simulations, we compute those functions for the Bethe lattice and the 4D Edwards-Anderson model. The details of the computation of the probability distributions $P^N(q)$, $P^N(q;h)$, and $P^N_C(q;h)$ for different system sizes $N$ in the simulations can be found in \textit{Materials and methods}.

\begin{figure}[t]
\centering
\includegraphics[width=1.\linewidth]{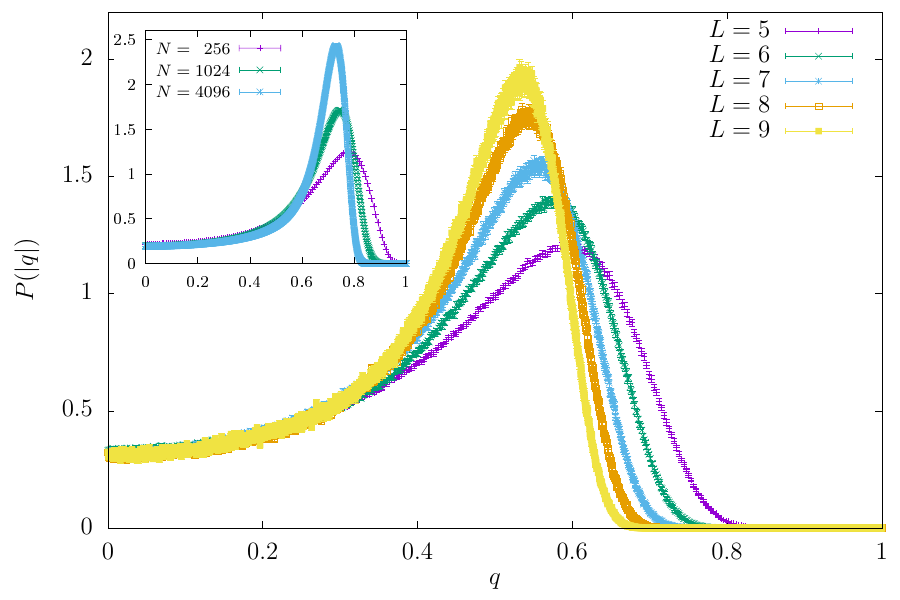}
 \caption{The probability $P^N(|q|)$ at $h=0$ for different values of $L$ in 4D EA at $T=0.7\; T_c$, $T_c(h=0)=2.03(3)$ \cite{marinari:99b}. Notice the stable part at low $q$ and the size-dependent peak, whose position shifts to the left, while its height and width respectively increase and decrease as the system size grows. 
\textbf{Inset:} The probability $P^N(|q|)$ at $h=0$ for different values of $N$ in Bethe lattice at $T=0.5\; T_c$, with $T_c(h=0)\simeq 1.518651$.}
\label{fig:PQ}
\end{figure}

The probability distributions $P^N(q)$ ---as well as $P^N(q;h)$ and $P_C^N(q;h)$--- have a strong dependence on $N$ (see Fig.~\ref{fig:PQ}), the height of the peak increases (it should go to infinity), and the position of the peak slightly drifts. Now the extrapolation to $N\to \infty$ is not easy and the results depend on the  functional form of the finite volume corrections. We avoid this problem by the following trick: we suppose that  the infinite volume $P(q)$ is well approximated by $P^N(q)$ and we compute $P^N(q;h)$ and $P_C^N(q;h)$ under this hypothesis.
To minimize finite volume effects, instead of comparing directly the overlap probability distributions, we found it convenient to consider, for $|q| \leq q_{\mathrm{EA}}$, the ratio of the probabilities distributions of the overlaps in the presence and absence of the field 
\begin{equation}
R^N_C(|q|;h)=\frac{P^N_C(|q|;h)}{P^N(|q|;0)}\,,\quad
R^N(q;h)=\frac{P^N(q;h)}{P^N(q;0)}\,.
\end{equation}

A crucial prediction of RSB is that these functions $R$   remain nontrivial in the infinite volume\footnote{Notice that in the limit of $T\to 0$ $R_C(q_\mathrm{EA},h)$ represents the \'survival\' probability that the ground state has not crossed in the interval of fields $[0,h]$. The expression we write here, provides a natural finite temperature generalization of this survival probability.} and are equal to the one computed analytically with the procedure described above. We can compare the numerical and the theoretical values of these functions, or compare for example  $P^N_C(|q|;h)$ with  $P^N_C(|q|;h)$ multiplied by the analytic prediction for $R^N_C(|q|;h)$ obtained by RSB theory.

\subsection*{Results}
\begin{figure}[t]
\includegraphics[width=1.0\linewidth]{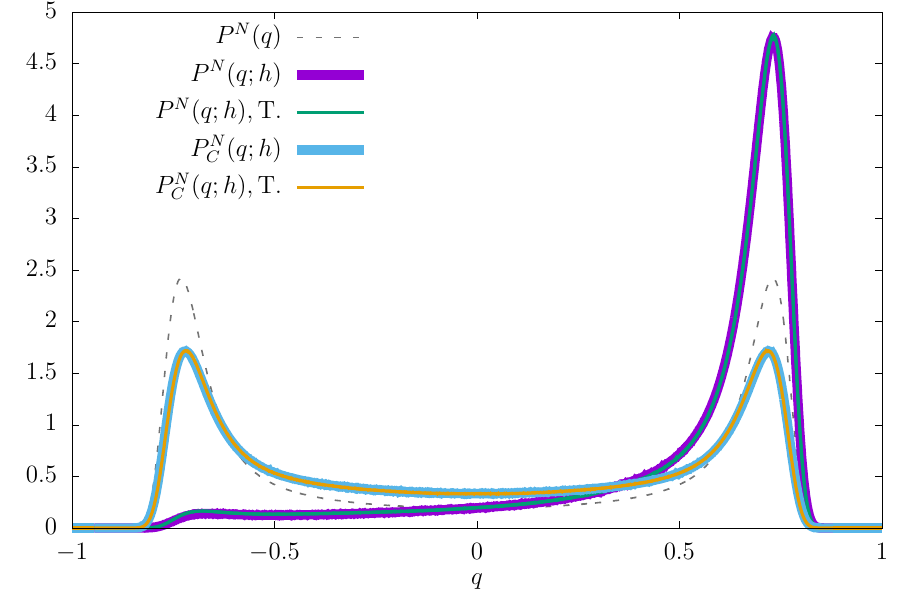}
 \caption{The functions  $P^N(q)$,  $P_C^N(q;h)$ and  $P^N(q;h)$ as a function of $q$ at $T=0.5 \; T_c$ for $h=4$  on Bethe Lattice for the largest volume simulated $N=4096$. We compare both the $P_C^N(q;h)$ and  $P^N(q;h)$ curves with the predictions from the ultrametric tree (T.). The dashed line is the zero-field $P^N(q)$ function that is the input in our computation. Notice that the probability distributions $P^N(q)$, and $P_C^N(q;h)$ are fully symmetric. Therefore, in the main text, we focus the discussion on $P^N(|q|)$, and $P_C^N(|q|;h)$. The curves  $P_C^N(q;h)$ and  $P^N(q;h)$ have been obtained as the product of the theoretical $R^N(q;h)$ and $R_C^N(q;h)$ with the numerical $P^N(q)$.
 \label{fig:Bethe}}
\end{figure}

\begin{figure}[h]
\includegraphics[width=1.0\linewidth]{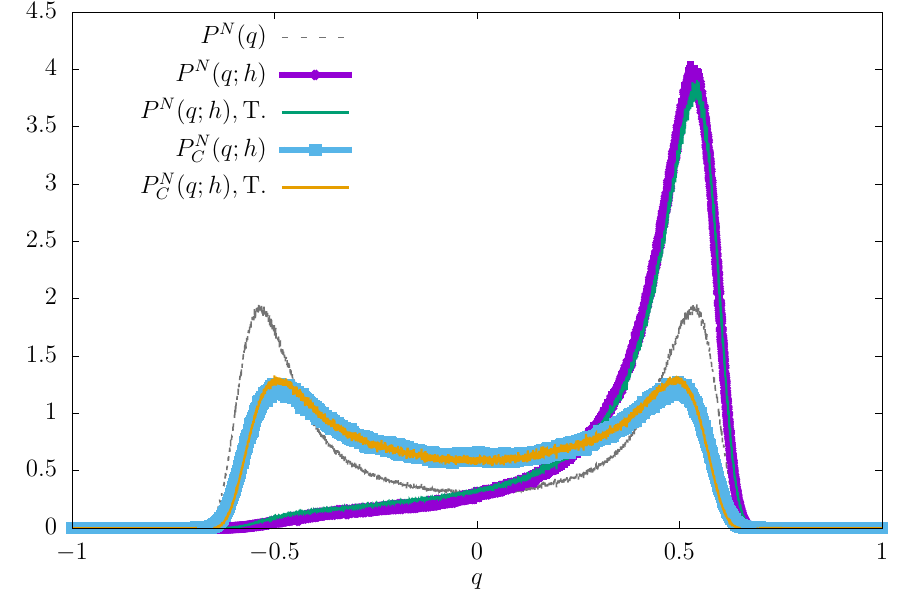}
 \caption{The functions  $P^N(q)$, $P_C^N(q;h)$ and $P^N(q;h)$ as a function of $q$ at $T=0.7 \; T_c$ for $h=10$  on the 4D Edwards Anderson for the largest volume simulated $N=9^4$. We compare both the $P_C^N(q;h)$ and  $P^N(q;h)$ curves with the predictions from the ultrametric tree (T.). The dashed line is the zero-field $P^N(q)$ function that is the input in our computation. Notice that the probability distributions $P^N(q)$, and $P_C^N(q;h)$ are fully symmetric. Therefore, in the main text, we focus the discussion on $P^N(|q|)$, and $P_C^N(|q|;h)$. The curves  $P_C^N(q;h)$ and  $P^N(q;h)$ have been obtained as the product of the theoretical $R^N(q;h)$ and $R_C^N(q;h)$ with the numerical $P^N(q)$.
 \label{fig:4D}}
\end{figure}

The numerical and theoretical probability distributions are shown in Fig. \ref{fig:Bethe} for the Bethe lattice and in Fig. \ref{fig:4D}  for the 4D EA model. In both cases, we find an excellent agreement between the theoretical predictions from the ultrametric tree and the numerical results for the functions  $P_C^N(q;h)$ and  $P^N(q;h)$. We also plot the numerical $P^N(q)$, which is used as an input for the theoretical prediction.

In these two figures, we show the results for the largest system we have simulated and the reader may be wondering what happens for smaller systems (because the functions $P_N(q)$ vary with $N$).

In order to appreciate better the quality of the comparison we find it more convenient to plot the analytic and numerical ratios  $R_C^N(|q|;h)$ and $R^N(|q|;h)$. In this way, we can better appreciate the deviations from the theory in the region where the functions $P$'s are small. The detailed results for the function $P_C^N(|q|;h)$ and for $P^N(q;h)$ are shown in the SI.

\begin{figure}[t]
\includegraphics[width=1.0\linewidth]{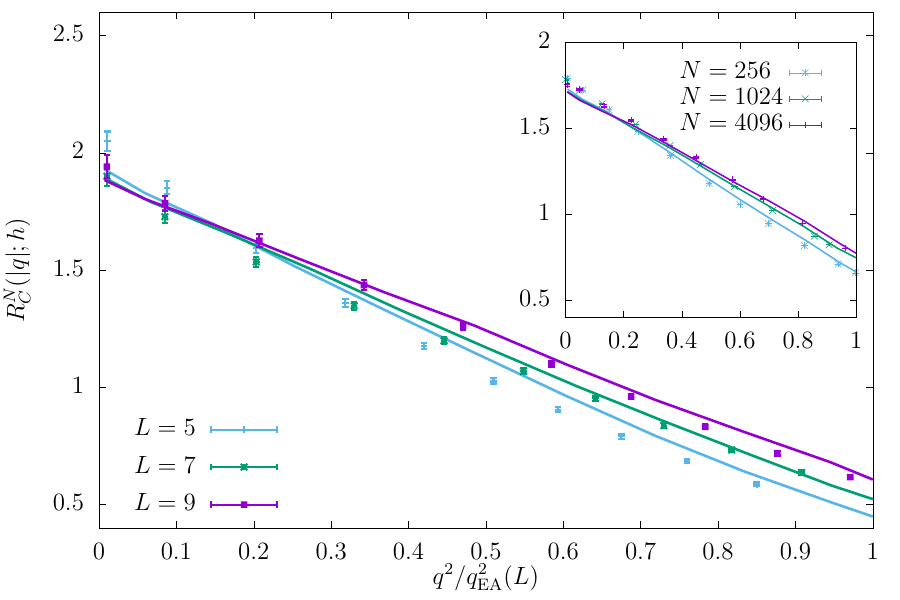}
 \caption{The ratio $R_C^N(|q|;h)$ as a function of $q^2/q_\mathrm{EA}^2(L)$ at $T=0.7 \; T_c$ for $h=10$ and different values of $L$ in the 4D EA model. Lines correspond to the theoretical prediction from the ultrametric tree and points correspond to simulations. For visualization purposes, we have plotted only a discrete subset of values of $q$. Inset: Ratio $R_C^N(|q|;h)$ as a function of $q^2/q_\mathrm{EA}^2(N)$ at $T=0.7 \; T_c$ for $h=4$ and different values of $N$ in the Bethe lattice. 
 \label{fig:PRC}}
\end{figure} 

In Fig. \ref{fig:PRC} we compare the theoretical predictions for $R_C^N(|q|;h)$ with the 4D EA computations, and also with the Bethe lattice ones (in the inset).  Interestingly, the ratio is  a linear function of  $q^2$ with a good approximation. The plot is done in the region $|q|<q_\mathrm{EA}$ where we can directly compare the data with the analytic predictions. The statistical errors increase in the region of small $|q|$ because in that region the probability $P(q)$ is smaller than in the peak. 
\begin{figure}[t]
\centering
\includegraphics[width=1.0\linewidth]{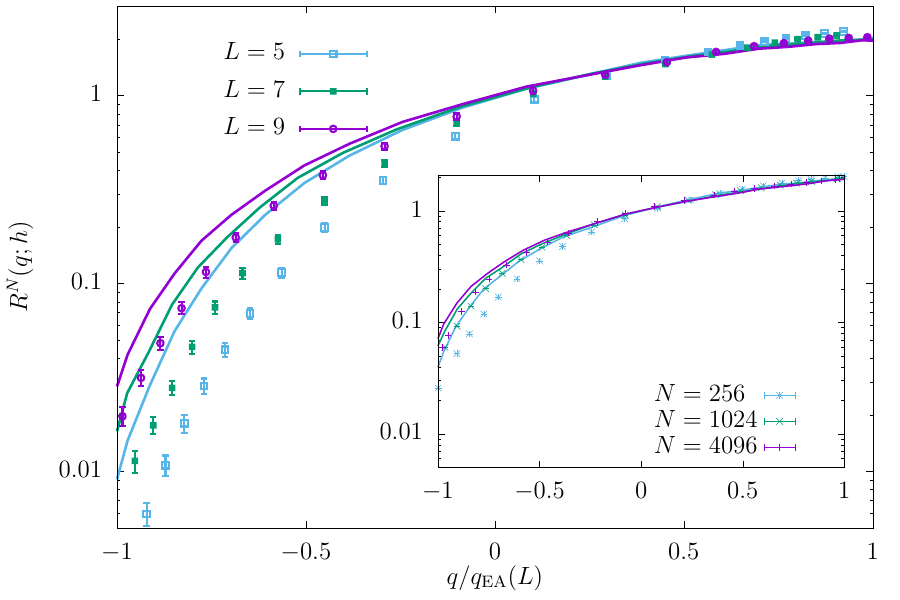}
 \caption{The ratio $R^N(q;h)$ as a function of $q/q_\mathrm{EA}(L)$ at $T=0.7\; T_c$ for different values of $L$ in 4D EA lattice ($h=10$). Lines correspond to the theoretical prediction from the ultrametric tree and points correspond to simulations. For visualization purposes, we have plotted only a discrete subset of values of $q$. Inset: ratio $R^N(q;h)$ as a function of $q/q_\mathrm{EA}(N)$ at $T=0.5\; T_c$ for different values of $N$ in Bethe lattice ($h=4$). 
 \label{fig:PRH}}
\end{figure}
 Furthermore, in Fig. \ref{fig:PRH}, we compare the theoretical predictions for $R^N(|q|;h)$ with the 4D EA computations and with the Bethe lattice ones (in the inset). Again, the plot is restricted to the region $|q|<q_\mathrm{EA}$ where we can directly compare the data with the analytic predictions. The data at negative $q$ have stronger volume dependence. The reasons are quite clear. The theory predicts that this ratio is 1 at $h=0$ because by continuity in $H$,
 $\mathrm{Prob}(|q|;h)\equiv P(q;h)+P(-q;h)$ is independent of $h$\footnote{This can be seen as a consequence of the invariance of the probability of the weight's distribution under the re-weighting due to the field, as implied by stochastic stability; indeed also in our analytic computations the property is exactly valid only in a limit, namely $M\to\infty$ see {\bf Methods}.}.
  On the other hand, for any {\it finite} field $H$, 
  the negative values of $q$ should have zero probability in  
  the infinite volume limit.

 Therefore the region of $q$ near zero (and consequently also for $q<0$) is strongly affected by finite size effects.

 We see that for small sizes the analytic computations are less accurate but qualitatively correct: this should be expected as far as the analytic computations are exact only in the infinite volume limit. Moreover, the peaks are wider (see Fig.~\ref{fig:PQ}) and the tail of the peaks sometimes arrives near zero: in this situation, our fitting procedure for the function $P(q)$ has a higher degree of approximation. The important message is that the difference between theory and numerical data strongly decreases by increasing the volume.

The asymmetry of the function $P^N(q;h)$ is evident (see Fig.~\ref{fig:Bethe} and Fig.~\ref{fig:4D}): remember that in the limit $h\to\infty$, $P^N(q;h)$ for $q<0$ must be zero, and $P^N(q;h)=2P^N(q)$ for $q\ge 0$, as it happens at an arbitrary small non zero value of $H$ in the infinite volume limit. For finite $h$ the two curves $P^N(q;h)$ and $P^N(q)$ cross at $q=0$. The chaos effect can be clearly identified as a decrease of the two peaks of $P_C^N(q;h)$ at 
$q=\pm q_\mathrm{EA}$ and by an increase in the region somewhat far from the peak, in particular in the region of $q$ near zero. Asymptotically, for large $|h|$, $P_C^N(q;h)$ should become a delta function $\delta(q)$ and we are very far away from this limit.

 Finally, an important observable, that could be directly measured in experiments, is the average magnetization as a function of $h$ that verifies: 
\begin{equation}
m(h)=\beta h \left(1-\int_{-1}^1 \dd q \;P(q;h)\; q \right)\,.
\end{equation}
as shown in the SI.

\begin{figure}[t]
\centering
\includegraphics[width=\linewidth]{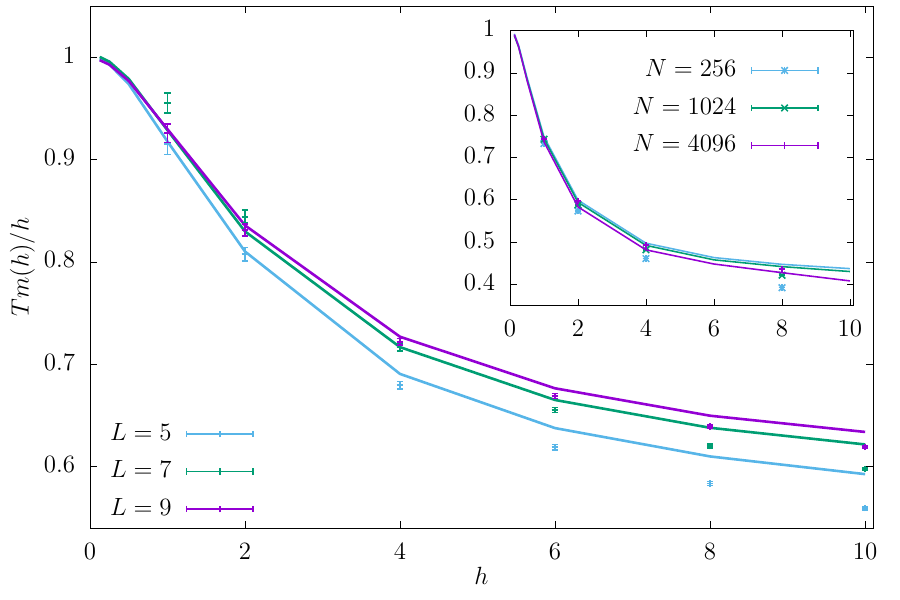}
 \caption{$T m(h)/h$  as a function of $h$ for different values of $L$ in the 4D EA lattice against the analytical prediction (continuous line) at $T=0.7T_c$. Inset:  $T m(h)/h$  as a function of $h$ for different values of $N$ in the Bethe lattice at $T=0.5T_c$. } 
 \label{fig:M}
\end{figure}
The results are presented in Fig. \ref{fig:M} which displays an excellent agreement between theory and simulations in both our simulated systems.   Let's discuss the implication of this result for the magnetization as a function of the standard field $H$: 
\begin{itemize}
\item For a very very small  magnetic field ($H\ll N^{-1/2}$), the magnetization  is trivially given by $\beta H$.
\item When $HN^{1/2}$ is of order 1, we enter the {\sl first} non-linear regime that is analyzed in this paper and depicted in  Fig. \ref{fig:M}.
\item When $H$ is small but $HN^{1/2}$ is large 
the magnetization per spin is given by the standard linear response formula $\beta H \left(1-\int_{0}^1 \dd q \;P(q)\; q \right)$ that is well understood, especially in solvable problems such as the SK model.
\item When $H$ is finite we enter the second non-linear regime where the thermodynamic functions depend on the magnetic field.
\end{itemize}   

Analogously to what we have seen in the ferromagnetic case,
the approach of the present
paper does not give 
information on the second non-linear regime.
We work in the regime where the function $P_N(|q|,h)$ does not depend on $h$ and, 
for example, we 
can not get any information on the existence of a De Almeida-Thouless line and an RSB phase at a nonzero -finite- value of the magnetic field. 
\section*{Discussion}
We have seen that RSB theory provides universal predictions on the behavior of the system in the presence of a small random perturbation, in our case a magnetic field, that only depends on the overlap statistics at zero field. We have tested these predictions through numerical simulations of relatively small spin glass systems. The Bethe-lattice spin glass gave us an idea of how finite volume corrections modify the asymptotic results. Remarkably a very similar behavior was found in the 4D Edwards-Anderson model. This method provides further quantitative tests of the RSB predictions in this system which adds to the one of Ref \cite{parisi1996equilibrium} where one finds results in agreement with the generalized fluctuation-dissipation relations. We have seen that in the crossover region replica symmetry breaking provides zero parameters quantitative predictions of functions of two variables ---the probabilities $P(q;h)$--- using as a starting point only the function $P(q)$ at zero magnetic field. The excellent agreement with numerical data further hints at the correctness of the standard RSB theory in this 4D model. On the contrary, a serious disagreement between the numerical simulations and the analytic results would have falsified the correctness of the standard RSB.

Our method could have experimental relevance in laboratory spin glasses. Quantitatively testing RSB theory experimentally is notoriously difficult. In Ref. \cite{franz1998measuring} it has been argued that, under a suitable hypothesis of stability of the distribution of states against small but extensive perturbations (stochastic stability), the measure \cite{herisson:02} of the modified fluctuation-dissipation relations in out of equilibrium ageing dynamics \cite{cugliandolo:93}
would directly yield access to the function $P(q)$. From this function, one could get a parameter-free prediction for the average magnetization in a small field $m(h)$.
The same quantity could be obtained in direct separate measurements in well-thermalized mesoscopic samples using small-size spin-glass powders. The comparison of the measured values of the function $m(h)$ with the theoretical predictions would provide a strong consistency check of the hypothesis of Ref. \cite{franz1998measuring} and those of RSB theory. The interesting regime is in the {\sl first  } nonlinear regime, i.e. at fields smaller than those where the standard linear response regime works. 

Chaos in a magnetic field is also an excellent alternative to temperature changes for studying memory and rejuvenation effects.  
This is particularly interesting because one can develop more precise theories than in the case of temperature changes.

\section*{Methods}
\subsection*{The tree of spin glasses}
Let us now briefly describe how to generate RSB spin-glass trees of states with the right statistics. 
In spin glasses, the equilibrium measure of a given infinite volume instance is characterized by a random weighted tree of states. 
A tree is completely characterized by its branching points, by a set of random weights $w_\alpha$ of the states that are associated to the leaves $\alpha$, and by the mutual overlaps $q_{\alpha\gamma}$, with $q_{\alpha\alpha}=q_\mathrm{EA}$. Since such trees have been discussed many times \cite{mezard:84b,mezard:87}, we will limit ourselves to a short description rather than to a full explanation. 

\begin{figure}[t]
\centering
\includegraphics[width=0.95\linewidth]{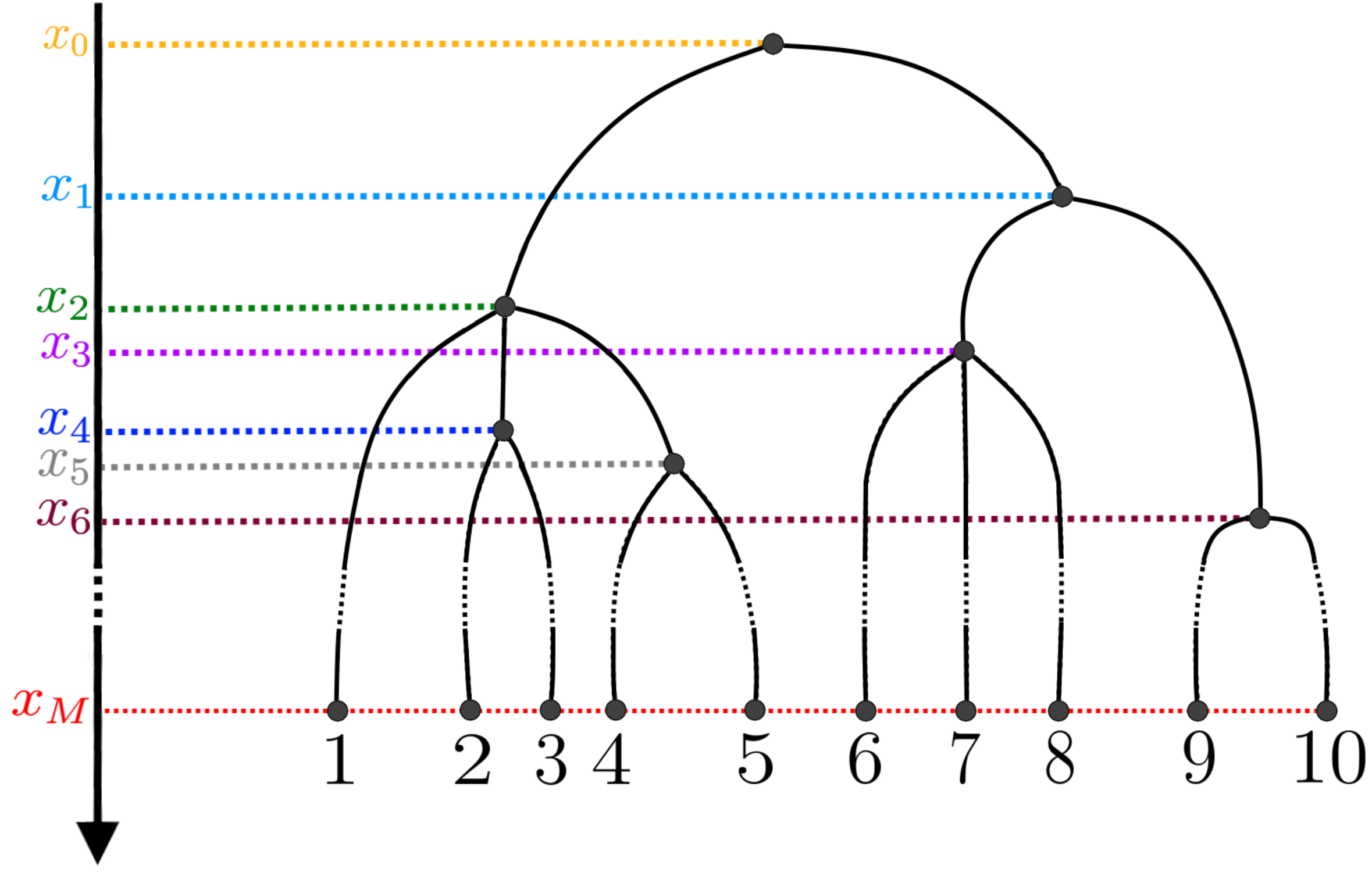}
\caption{A tree with ten leaves. The overlap of leave (1) with the leaves ($2\cdots5$) is $q(x_2)$ and with the leaves ($6\cdots 10$) is $q(x_0)$; the overlaps of leave (2) with the other leaves are the same as leave (1), with the only difference that the overlap with the leave (3) is now $q(x_4)$.}
\label{fig:tree}
\end{figure}

In principle, trees have an infinite number of leaves, but by cutting off small weights, we can approximate them by trees with a finite number of leaves $M$.  See, for example, Fig. \ref{fig:tree}.

These can be generated through a branch merging process known as Bolthausen-Sznitman coalescent \cite{bolthausen:98}, starting from $M$ leaves, and progressively reducing the number of branches through a Markov process of random collisions in a "time" $t$. A direct branching algorithm can also be used~\cite{parisi2015explicit}, however, here we find it more convenient to use the coalescent\footnote{We have run codes with both prescriptions, with identical results, but the Bolthausen-Sznitman code is simpler and more efficient.}.

The branching points of the tree are labeled by the time variable $t$ ($t\ge 0$).
In the following, we call nodes both the branching points and the leaves (terminal nodes). The coalescence process is ruled by the function $P(k|b)=\frac{b}{(b-1) (k-1) k}$ which tells the probability of coalescence of $k$ nodes if at that time there are $b$ nodes. 
We start with $M$ nodes at time zero: in this case $b=M$.  We chose a number $k\in \{2,...,b\}$ at random with probability $P(k|b)$  and we coalesce $k$ nodes at random into a single one. In this way $b$ decrease to $b-k+1$. At the same moment, the time variable is incremented by a random $\Delta t$ 
with an exponential distribution and with average $ \overline{\Delta t}=1/(b-1)$. The process stops at a finite time when $b=1$ and no more coalescence is possible.

At each level, $t$ is associated with a value $x(t)=x_M \exp (-t)$:  the overlap between the branches that meet at that level is given by $q=q(x(t))$. 
This is the only point where the function $P(q)$ appears in the construction. In our case, we obtain this $P(q)$ from the numerical data (see below for further details). Once the tree is built, one needs to associate random weights to the leaves. If $M$ is large, this is simply done by generating $M$ i.i.d. variables $u_\alpha$, for $\alpha=1,...,M$ with distribution $p(u)= x_{M} \theta(u-1) u^{-1-x_{M}}$, and defining the weights $w_\alpha=u_\alpha/\sum_\gamma u_\gamma$. Notice that in this construction both the weights and the overlaps  associated with the given branching levels are independent of the tree's structure. The values of $x_M$ and of $x(q)$ depend on the specific system at hand and should be given as input at the beginning of the computation.

The magnetizations $m_\alpha$ are zero mean Gaussian random variables with covariance $\overline{m_\alpha m_\beta}=q_{\alpha\beta}$ and they can be constructed efficiently in many ways. More details are provided in the SI.

\subsection*{Comparison with numerical simulations} 
We simulate  Ising spin glass models at equilibrium in their low-temperature phase.
The Hamiltonian of a system with $N$ spins in a field reads 
\begin{equation}
    {\cal H}=-\sum_{(i,j)\in {\cal E}} J_{ij} S_i S_j -\frac{h}{\sqrt{N}}\sum_{i=1}^N S_i \label{eq:hamiltonian}
\end{equation}
Where $\cal E$ denotes the set of edges of the graph where the model is defined:
in our case either a Bethe lattice spin glass on a Random Regular Graph with coordination number equal to 4, or the 4D Edwards-Anderson model with periodic boundary conditions in a hypercubic lattice. 

In a finite volume, the function $P(q)$ is measured as the probability distribution of the overlap between two equilibrium configurations (the replicas denoted as $\sigma$ and $\tau$) $q=\sum_i\sigma_i \tau_i/N$. We measure the probability $P(q)$ at $h=0$ and the probabilities $P_C(|q|;h)$ and $P(q;h)$ as a function of $h$ at low temperatures. In both cases, optimized code based on parallel tempering and multispin coding with 128 bits has been used. In Fig. \ref{fig:PQ} we show the results for the probability $P(|q|)$ for both systems at zero magnetic field. 

From our numerical data, we extrapolate the functions $x(q)$ from these probability distributions at infinite volume. For each simulated volume, we approximate $P(q)$ with  $P_\mathrm{fit}(q)=a+b q^2+(1-x_M)\delta (q-q_\mathrm{EA})$: so, $x(q)=aq +bq^3/3 $ and  $x_M=x(q_\mathrm{EA})$.  Then we fit the values of $a$ and $b$ using the data in the low $q^2$ region and we fix $q_\mathrm{EA}$  such that the mean value $\overline{\langle |q|\rangle}$ computed with $P_\mathrm{fit}(q)$ coincides with that obtained in the numerical simulation for that particular volume. In this way,  we find that the position of the peak at that given volume is only a few percent off the estimated $q_\mathrm{EA}$ (see the SI). Using this representation $P_\mathrm{fit}(q)$ of the function $P(q)$ we have all the information to assign the weights of the leaves in the building process of the tree (see above). The predictions from the random trees have been averaged over $10^5$ trees with $10^5$ leaves: all the results are stable toward a variation of the number of leaves and the statistical errors are small.
 We concentrated our attention on $\overline{\langle |q| \rangle}$ because we know that for large values of $h$ the magnetic susceptibility depends linearly on it (see the SI).

The reader may note that the theoretical curves of the probability distributions appearing in Fig.~\ref{fig:Bethe} and Fig.~\ref{fig:4D} are labeled with the superindex $N$ which, in the theoretical case, is not associated with any size of the system. This superindex refers to the numerical curve (associated to a finite lattice with $N$ spins) from which the $P_\mathrm{fit}(q)$ is obtained following the above-explained process. In particular, the value $x_M$ used as the input in the construction of the tree depends on $N$ and this dependence is inherited by the tree.

An alternative approach would be to extrapolate the function $P(q)$ to infinite volume and to use this extrapolated function for the analytic computation.
Using no extrapolation is more robust and allows us to make predictions even 
for a single value of $N$.

There is an asymmetry between our analytical approach and the numerical simulations:
\begin{itemize}
    \item In the analytic approach, we generate the weights of the states at $h=0$ and we reweight them by adding a magnetic field. The reweighting factor may be quite large: it is a number of order $\exp \left( \beta h q_\mathrm{EA}^{1/2}z \right)$ where $z$ is a Gaussian number of unit variance and zero average. Therefore states that have a small probability at $h=0$ may become dominant at $h\ne 0$. For that reason, we have to compute states with a very small probability and this can be done by considering a sufficiently large number of states ($M$).
    \item In the numerical case, we simulate the systems both at zero and  non-zero magnetic field. The reader may wonder why we do not use  the Swendsen-Ferrenberg reweighting method for computing numerically the probability distributions in the presence of the field just reweighting the configurations generated by simulations at zero field. This technique works for small values of the magnetic field, however, in our case, the interesting region is for relatively large field $h$ (e.g., 4) where the Ferrenberg-Swendsen~\cite{ferrenberg:88} reweighting method fails unless the simulation length becomes very large.
\end{itemize}

\subsection*{Data, materials and Software Availability}
 The data and the scripts that generate the figures of this article can be downloaded from \url{https://github.com/maguilarjanita/SmallFieldChaos}~\cite{github_rep}.

\begin{acknowledgments}
We acknowledge the support of the Simons Foundation (grants No. 454941, S. Franz and No. 454949, G. Parisi). We were partly supported as well by Grants No. PID2022-136374NB-C21 and PID2020-112936GB-I00, both of them funded by MCIN/AEI/10.13039/501100011033, FEDER, UE.  This research has been supported by ICSC – Italian Research Center on High Performance Computing, Big Data and Quantum Computing, funded by the European Union – NextGenerationEU. MAJ was supported by Community of Madrid and Rey Juan Carlos University through Young Researchers program in R\&D (Grant CCASSE M2737). JMG was supported by the Ministerio de Universidades and the European Union NextGeneration EU/PRTR through 2021-2023 Margarita Salas grant.
\end{acknowledgments}

\bibliography{biblio}
\newpage
\appendix

\textbf{\begin{center}
  \bf SUPPORTING INFORMATION
\end{center}}

\section*{SI: Some technical details on the analytic computation}

In the RSB theory in the infinite volume limit, each sample, i.e. each choice of the couplings $J$, is characterized by the weight of the states, their mutual overlap, and their magnetizations. These quantities fluctuate from sample to sample, the theory therefore deals with the probability distribution of these quantities. Ultrametricity implies that for each sample the states can be considered as the leaves of a tree, where the leaves carry a weight. Our task here is to generate numerically the weighted trees with the correct probability distribution. 

In the real world quantities like the $P_J(q)$ depend on the system via the couplings $Js$: in the analytic approach one determines the probability distribution of the probability distribution $P_J(q)$. The simplest way to determine the probability of probabilities is by describing an algorithm that generates it. To this end, we show how to compute a $P_\mathrm{Tree}(q)$ that depends on the tree in such a way that in the large volume limit, $P_J(q)$ and $P_{\mathrm{Tree}}(q)$ have the same statistical properties.

Through random trees, we can approximate the distribution of spin-glass Gibbs measure in the thermodynamic limit.  For the purposes of this  work, we need for a given sample:
\begin{itemize}
    \item A set of $M$ weights $w_\alpha$ for each state $\alpha=1,...,M$.
   \item A matrix of the overlaps $q_{\alpha\gamma}$.
      \item A set of Gaussian magnetizations $m_\alpha$ with covariances $\overline{m_\alpha m_\gamma}=q_{\alpha\gamma}$.
\end{itemize}

The first difficulty is that in the theory the number of leaves of the states is infinity. However, 

if we replace infinity by a large number we only commit a small error. 
If we neglect the leaves that have a weight less than $\epsilon$, the pruned tree remains with a finite number $M$ of leaves ($M= O(\epsilon^{-x_M})$). Moreover, in the limit where the cutoff $\epsilon$ goes zero (or $M$ goes to infinity), the interesting quantities averaged over the weighted pruned trees reproduce the same quantities averaged over the infinite weighted trees.

In principle, it is easy to 

generate trees according to the algorithm described in the text.
However, it is not immediate to write an efficient code. There are many possibilities. Let us describe our choices.
The random weights are easily constructed:
\begin{equation}
 u_\alpha= \exp( x_M \log(r_\alpha)), \ \quad w_\alpha=\frac{ u_\alpha}{\sum_{\gamma=1}^M u_\gamma}\,, \label{eq:W}
 \end{equation}
 where $r_\alpha$ are i.i.d. random numbers that are uniformly distributed inside the interval $[0,1)$. Notice that the minimum weight $w_{\min}$ is of order $\epsilon_M \sim M^{-1/x_M}$.  Depending on the value of $x_M$ and the perturbation, the value of $M$ may be adequate (or not).
 
We then build up the tree with the aid of two variable length lists: 
\begin{itemize}
    \item The list of the identifiers (IDs) of the active nodes contains the nodes that have not yet coalesced and do not yet have a parent. 
    \item The list of all nodes present at a given moment. Each node, indexed by its ID, is represented by a structure that contains the time at which it has been created and a variable $v$ conventionally equal to $-1$ if the node is active and equal to the ID of the parent node if the node has coalesced.   
\end{itemize}
In addition, we have the time variable $t$. Let's call $M(t)$ and $b(t)$ respectively the total number of nodes and the number of active nodes at time $t$. 
Clearly, $M(0)=b(0)=M$, while for large $b(t)$ we have approximately $b(t)\simeq M^{e^{-t}}$. 

We repeat the following procedure up to the moment $t_\mathrm{fin}$, when $b(t_\mathrm{fin})=1$, i.e. there is only one surviving node, so no further coalescence is possible. 
\begin{itemize}
    \item We increment the time variable $t$  by a random $\Delta t$ with an exponential distribution and average $1/(b(t)-1)$: $\Delta t=- \log(r)/(b(t)-1)$ where $r$ is a random number uniformly distributed inside the interval $[0,1)$.
    \item We chose a number $k\in \{2,...,b(t)\}$ at random with probability $P(k|b(t))=\frac{b(t)}{(b(t)-1)(k-1)k}$ and we coalesce $k$ nodes chosen at random into a single one. In this way $b$ decreases to $b-k+1$ (i.e. $b(t+\Delta t)=b(t)-k+1$).
    \item We thus add a new node to both lists of nodes: its timestamp is set equal to $t+\Delta t$, its variable $v$ is set to -1 and its ID is equal to the last ID created incremented by one. 
    \item We add to the $k$ branching nodes the information on the parent, and we remove them from the list of active nodes. 
    \item We stop when only a single node is active.  
\end{itemize}
When we stop only the root of the tree has an ID of the parent equal to $-1$. In this construction, the timestamp of the node increases with the ID number. 
We would like to stress that up to this point the construction of the tree is universal, i.e. it is the same for any value of $x_M$ (if we neglect the weights) and any function $q(x)$. 
The construction of the magnetizations can be done as follows, 
scanning the tree from the root to the leaves:
\begin{itemize}
    \item To each node, we assign an overlap given by $q(x_M \exp(-t))$.
    \item We start with the root and we assign to it a root magnetization that is equal to $r_g \sqrt{q_{\text{root}}}$, where $r_g$ is a Gaussian random number with variance one. 
    \item We now scan the list of the nodes and we set
\begin{equation}
m_{\text{node}}= m_{\text{parent}}+r_g \sqrt{q_{\text{node}}-q_{\text{parent}}}\,.
\end{equation}
\item In this way, the magnetization of the leaves are Gaussian random variables with the required covariance, 
$\overline{m_\alpha m_\gamma}=q_{\alpha\gamma}$, 
as it can be readily checked.

\end{itemize}
 It is in principle straightforward to compute quantities like:
 \begin{equation}
 P_{J,C}(q;h)=\sum_{\alpha,\gamma} w_\alpha w_\gamma(h) \delta (q- q_{\alpha\gamma})\,.
 \end{equation}
 However, the sum contains $M^2$ terms, which is computationally heavy for large $M$. 
 
 A fast approximate computation consists of the following. Let $\epsilon_M$ be the minimum value of $w_\alpha$ ($\epsilon_M= O(M^{-1/x}))$. We can neglect pairs with $ w_\alpha w_\gamma(h) <\epsilon_M$. The number of the surviving terms is of order $O(M\log M)$ and we can restrict the sum 
 to these terms if we order both the $w_\alpha$ and the $w_\gamma(h)$ in decreasing order and sum over $\alpha$ and $\gamma$ using two nested loops. 
 
A final warning: most of the computer time may be spent in the computation of the $O(M^2)$ overlaps. There are many possible ways to do this computation and one should be careful in using an efficient one that exploits the tree structure of the states.

In the end, the computational time can 
be reduced from a naive $O(M^2)$ to  $O(M (\log M)^z)$. The exponent $z$ is likely equal to 2, but we have not measured the precise timings.

\subsection*{Average Magnetization}

Let us study the average magnetization $m(h)$ in the presence of the field $h$. As in the main text, we define $m$ as the total magnetization divided by $N^{1/2}$ in such a way that for large $h$ the quantity $m(h)/h$ becomes the susceptibility. In principle, two phenomena contribute to the average magnetization:
\begin{itemize}
 \item  Within each state, the usual fluctuation-dissipation theorem holds, and the magnetic field increases the average magnetization by an amount of $\beta h(1-\overline{m_\alpha^2})=\beta h(1-q_\mathrm{EA})$ 
 \item  As we have discussed at length, states with higher average magnetization $m_\alpha$ become more likely in the presence of the field $w_\alpha\to w_\alpha(h)$.
\end{itemize}
Taking care of both contributions we find
\begin{equation}
    {m(h)}= \beta h (1-q_\mathrm{EA})+\overline{\sum_\alpha w_\alpha(h) \tanh (\beta h m_\alpha)m_\alpha}\,.
\end{equation}
The second term in the previous equation can be readily computed using the same strategy as before by averaging over randomly generated trees. Moreover, integration by parts over the Gaussian distribution of the $m_\alpha$ reveals that the last term equals to

\begin{equation}
\begin{split}
&\overline{\sum_\alpha w_\alpha(h) \tanh (\beta h m_\alpha)m_\alpha}=\\
&\beta h \bigg(q_\mathrm{EA}-\overline{\sum_{\alpha\beta}w_\alpha(h)w_\beta(h)\tanh(\beta h m_\alpha)\tanh(\beta h m_\beta)q_{\alpha\beta}}\bigg)=\\
&\beta h \big(q_\mathrm{EA}-\langle q\rangle_h\big)\,.
\end{split}
\end{equation}
Putting the two terms together, we finally find
\begin{equation}
m(h)=\beta h \left(1-\int_{-1}^1  q \, q P(q;h) \right)=\beta h \big(1 -\langle q \rangle_h \big) \,.
\end{equation}
The quantity $m(h)/h$ is equal to $\beta $ at $h=0$ and goes to the thermodynamic susceptibility  $\chi=\beta\int_0^1 
\mathrm{d}x \;q(x)$ for large $h$. 

\section*{SI: Simulation details}

\subsection*{4D Edwards-Anderson} 
\label{subsec:simulation_EA4D}

We have studied the 4D EA model in the presence of an external magnetic field. Its Hamiltonian is given by  
\begin{equation}
    {\cal H}=-\sum_{(i,j)\in {\cal E}} J_{ij} S_i S_j -\frac{h}{\sqrt{N}}\sum_{i=1}^N S_i \label{eq:hamiltonian}
\end{equation}
where $N=L^4$ is the total number of spins living in a four-dimensional hypercubic lattice of size $L$ with periodic boundary conditions. The couplings $J_{ij}$ are drawn from a bimodal probability distribution: it can be $\pm 1$ with a 50 \% probability.

The disorder realization is quenched, i.e. the couplings remain constant through the whole simulation, defining what it is usually called a \textit{sample}. Moreover, we have simulated different \textit{replicas}, which are different simulations of the same sample evolving with different thermal noise.

Due to the toughness of the thermalization process in the simulation of spin glasses, it is convenient to speed up the convergence to the equilibrium by using different techniques. In our particular case, we have used a Multispin Coding Monte Carlo simulation~\cite{friedberg:70,jacobs:81} and we have taken advantage of the long binary registers that current CPUs can operate with: they allow us to simulate 128 samples at once. A set of 128 samples that are simulated together by using Multispin Coding is called \textit{a super-sample}. We have also performed a parallel tempering \cite{hukushima:96,marinari:98b} proposal every 20 Monte Carlo Sweeps. A detailed explanation of the implementation for the six-dimensional case in a magnetic field can be found in Ref. \cite{aguilar:23}.

We have simulated lattice sizes running from $L=5$ to $L=9$ with all the values of the external magnetic field in the set $h=\{0, 1, 2, 4, 6, 8, 10\}$. For each lattice size, we have simulated 20 $\times$ 128 samples.  The same set of  20 $\times$ 128 samples has been simulated for the seven different values of the external magnetic field given above. Moreover, for each sample, we have simulated two replicas. The number of temperatures of the parallel tempering depends on the system size, see Table \ref{tabla_sim}, but our work focuses on $T=0.7 T_\mathrm{c} \approx 1.421$ so this temperature is always the lowest one.

\begin{table}[h]
\centering
\begin{tabular}{c c c c c} \hline 
$L$ & ~~\#Samples~~~~& \#Temp. & ~$T_\mathrm{min}$ & ~$T_\mathrm{max}$ \\ 
\hline \hline
5 & 20 $\times$ 128 & 12  & 1.421    &2.800 \\ 
6 & 20 $\times$ 128 & 18  & 1.421    &2.800 \\ 
7 & 20 $\times$ 128 & 24  & 1.421    &2.800 \\ 
8 & 20 $\times$ 128 & 32  & 1.421    &2.800 \\ 
9 & 20 $\times$ 128 & 36  & 1.421    &2.800 \\ 
\hline \hline
\end{tabular}
\caption{Some parameters of our simulation. The first column refers to the linear size of the hypercube. The number of spins is $N=L^4$. In the second column, we present the number of samples analyzed, the specification $\times 128$ refers to the fact that each of the 20 super-samples contains 128 independent samples. The third column shows the number of temperatures simulated for each size. This number has been chosen in a way that ensures the random walk-in temperatures is sufficiently ergodic. Finally, the fourth and fifth columns refer to the lower and upper values of the temperature interval.}
\label{tabla_sim}
\end{table}

To ensure that we are working at equilibrium, the thermalization process must be monitored sample by sample.  In this work, we use the thermalization protocol originally appearing on Ref.~\cite{billoire:18} which we briefly explain here for convenience.

We first determine with preliminary runs a sufficiently large number of Monte Carlo steps for most of the samples to be thermalized, and then we simulate our 20 super-samples that number of Monte Carlo steps. Along the simulation, we record the random walk in the temperatures of the parallel tempering and we use this information to compute the integrated autocorrelation time $\tau_{\mathrm{int}, f}$ for several observables $f$, related to the random walks \cite{billoire:18}. We use now the largest value of those integrated autocorrelation times, $\tau_{\mathrm{int}, f*}$, to estimate the exponential autocorrelation time by assuming that $\tau_{\mathrm{int}, f*}\sim \tau_\mathrm{exp}$. Finally, we impose the criteria that a sample is considered thermalized when it has been running for a number of Monte Carlo steps thirty times bigger than $ \tau_\mathrm{exp}$.

It is possible that, inside a super-sample, all the samples are thermalized except a few ones. In that case, we take the last configuration of the non-equilibrated samples and extend the simulation as long as necessary to reach the above-exposed thermalization criteria.

\subsection*{Spin glasses on Bethe lattices} 
\label{subsec:simulation_Bethe}

To study spin glasses in a field on Bethe lattices we have considered the Hamiltonian in Eq.~\eqref{eq:hamiltonian} where $\mathcal{E}$ is the edge set of a random regular graph of fixed degree 4. The couplings $J_{ij}$ are drawn from a bimodal probability distribution: $J_{ij}=\pm 1$ with a 50 \% probability. For each sample, we have simulated 4 replicas without a field ($h=0$) and 4 replicas in a field, with several field intensities ($h\in\{1,2,4,8,16\}$).

We have sped up the simulation with the same techniques used in the 4D case, namely multi-spin coding and parallel tempering.
In the multi-spin coding, we have used words of 128 bits, which allows us to simulate in parallel 128 systems sharing the same interaction graph topology while having different random couplings. We then average over a large number of different random regular graphs of fixed degree 4: this number is the first one in the column ``\#Samples'' in Table~\ref{table_Bethe} and is reported with a range [min-max] since a different computational effort has been devoted to different values of $h$ (the largest number of samples is the one always corresponding to the values of $h$ reported in the Figures in the main text).

In the parallel tempering algorithm, we have performed an attempt to swap temperatures every 32 Monte Carlo sweeps. The temperature schedule has been optimized following the ``0.23 rule'' \cite{pelissetto2014large} and imposing $T_\text{min}=0.5 T_c$ and $T_\text{max}=1.5 T_c$, where $T_c=1/\text{arctanh}(1/\sqrt{3})\simeq 1.518651$ is the spin glass critical temperature. For the three sizes that we have simulated the number of temperatures used is reported in Table~\ref{table_Bethe} in the column ``\#Temp''. The column ``\#MCS'' in Table~\ref{table_Bethe} corresponds to the number of Monte Carlo sweeps we have run in each simulation. Typically the thermalization time (thanks to the use of the parallel tempering algorithm) is much smaller than that number. We have checked that the results obtained in the last half of the simulation are statistically equivalent to those obtained in the preceding quarter of the simulation. Data presented in the paper always corresponds to measurements obtained in the last half of the simulation. Moreover, the errors (when reported) have been always obtained only from graph-to-graph fluctuations: in this way, errors are somehow overestimated but are certainly insensitive to any correlation between measurements taken on the same sample and even on the same graph (with different couplings).

\begin{table}[h]
\centering
\begin{tabular}{c c c c} \hline 
$N$ & \#Samples & \#Temp. & \#MCS \\ 
\hline \hline
 256 & $[75005-144114] \times 128$ & 8   & 65536 \\ 
1024 & $[7614-11463] \times 128$  & 15  & 65536 \\ 
4096 & $1120 \times 128$          & 28  & 262144 \\ 
\hline \hline
\end{tabular}
\caption{Parameters of the simulations performed on Bethe lattices with fixed degree 4. A description of the parameters is provided in the text.}
\label{table_Bethe}
\end{table}

\section*{SI: Analysis details}
\subsection*{Computation of the $P(q)$} The basic quantity we are interested in is the overlap $q$. To compute this observable we need to know the spin field of two different replicas $\sigma$ and $\tau$ at equilibrium
\begin{equation}
q=\dfrac{1}{N} \sum_i \sigma_i \tau_i \, ,
\end{equation}
where the sum runs over all the spins of the system and $N$ is the total number of spins. 

Computationally, we take advantage of the Multispin Coding technique used to simulate our system.
Since our CPU can perform binary operations over 128-bit words and we have recorded configurations from 128 samples at the same place, we use this previous setup to compute packages of 128 overlaps at once.

The main observable of our work is the overlap probability distribution $P(q)$. First, we compute a  $P_J(q,L)$ for each one of the $2560$ samples and for each value of the field $h$. To do it we just take $10^4$ measures of the overlap $q$, as explained above, and build a histogram of frequencies. To avoid asymmetries induced by the binning process, we use one bin for each one of the $L^4+1$ possible different values of the overlap. Finally, we conveniently normalize the histogram to obtain a $P(q)$ such that $\int_{-1} ^{1} \mathrm{d}q~P(q)=1$. Once we have a $P_J(q)$ for each sample, we perform the average over the disorder and compute the error bars for each bin using the Jackknife method ~\cite{Jacknife_PY,efron1982jackknife}. 

Once we have computed the $P(q, L)$, we want to obtain the position of the peaks, $q_\mathrm{max}(L)$, for the $h=0$ case. For this case, RSB predicts that the two symmetric peaks observed at $L$ become two Dirac deltas at $q_\text{EA}$ in the thermodynamic limit. In this section, we consider only positive values of $q$ and the symmetrized version of $P(q)$. 

To obtain $q_{\mathrm{max}}(L)$  we begin by smoothing the $P(q)$ by taking its convolution with a Gaussian of width $1/\sqrt{8N}$. We define this smoothed version  $\mathcal{P}(q)$ as 
\begin{equation}
\mathcal{P}(q=c)=\int_{-\infty}^{\infty} \dd q^{\prime} P\left(q^{\prime}\right) \mathcal{G}_{8N}\left(c-q^{\prime}\right),
\end{equation}
where 
\begin{equation}
\mathcal{G}_{8N}(x)=\sqrt{\frac{8N}{2 \pi}} \mathrm{e}^{-8N x^2 / 2} .
\end{equation}
Now, working with the new  $\mathcal{P}(q)$, we fit the neighborhood of the peak to a third-order polynomial and define $q_{\text{max}}(L)$ as the maximum of this polynomial. We say that a given point belongs to the neighborhood of the peak if its height surpasses a value of $0.9$ times the maximum height of the  $\mathcal{P}(q)$. The values of $q_{max}(L)$ obtained by this procedure can be checked in Table ~\ref{tabla_qmax}. As explained in the \textit{Materials and methods} section of the main article, the position of the peak at a given volume is only a few percent of the estimated $q_\mathrm{EA}(L)$. In Table~\ref{tabla_qmax} we also include the values of $q_\mathrm{EA}(L)$ used for the computation of the ultrametric trees. 

\begin{table}[h]
\centering
\begin{tabular}{ccc} \hline \hline
$L$ & $q_\mathrm{max}(L)$ & $q_{\mathrm{EA}}(L)$ \\ \hline
5 &0.592(2) & 0.625\\
6 &0.574(2) & 0.585\\ 
7 &0.555(1) & 0.568\\ 
8 &0.543(1) & 0.542\\ 
9 &0.535(1) & 0.530\\ \hline \hline
\end{tabular}
\caption{Position of the peaks  $q_\mathrm{max}(L)$ of the $P(q,L)$  and values of $q_{\mathrm{EA}}(L)$ for different values of the lattice size $L$ in the 4D EA model for $T=0.7 T_\mathrm{c} \approx 1.421$.  }
\label{tabla_qmax}
\end{table}

\subsection*{Compuation of the $R$-ratios}
Finally, we will discuss how we have computed numerically the $R$-ratios.

In Fig.~\ref{fig:PEA}, we compare the theoretical results for $R_C^N(|q_\mathrm{EA}|)$ and $R^N(-|q_\mathrm{EA}|)$ with the data for those observables obtained from the numerical simulations on the 4D EA lattice. However, in finite systems, the function $P(q)$ does not end abruptly at $q=q_\mathrm{EA}$ but has a tail of non-zero probability up to $q=1$. Therefore, the definition of this observable in finite lattice systems is not obvious. In Fig.~\ref{fig:PEA}, we compute $R_c^N(|q_\mathrm{EA}|)$ as the integral from $q=q_\mathrm{EA}$ to $q=1$, and analogously, we compute $R^N(-q_\mathrm{EA})$  as the integral between $q=-1$ and $q=-q_\mathrm{EA}$. In particular: 

\begin{equation}
    R_C^N(|q_\mathrm{EA}|;h) = \frac{\int_{q_\mathrm{EA}}^{1} \dd q~P_C^N(q;h) }{\int_{q_\mathrm{EA}}^{1} \dd q~P^N(q;0)}\;,
\end{equation}
and 
\begin{equation}
    R^N(-q_\mathrm{EA};h) = \frac{\int_{-1}^{-q_\mathrm{EA}} \dd q~P_C^N(q;h)}{\int_{-1}^{-q_\mathrm{EA}} \dd q~ P^N(q;0)}\;.
\end{equation}
Similar results can be obtained if one defines $R_c^N(|q_\mathrm{EA}|)$ as an integral around the peak:
\begin{equation}
    R_C^N(|q_\mathrm{EA}|;h) = \frac{\int_{q_\mathrm{EA}-\Delta}^{q_\mathrm{EA}+\Delta} \dd q~ P_C^N(q;h)}{\int_{q_\mathrm{EA}-\Delta}^{q_\mathrm{EA}+\Delta} \dd q~ P^N(q;0)}\;,
\end{equation}
where we define the interval of integration $2\Delta$ as the maximum distance between the points that fulfill the condition of having a height greater than 0.8 times the height of the point at $q_\mathrm{EA}$. In fact, we use this definition of the observable to compute the results shown in Fig.~\ref{fig:LARGE}. 

In Figs.~4 and 5 of the main article we probe the functions $R_C$ and $R$ in the region $|q|<q_\mathrm{EA}$, for which we see that the dependence on the volume is rather mild.
In the same pictures, we plot the theoretical predictions that have been computed averaging over { $10^5$ trees with $M=10^5$ leaves}. 
{
Remarkably, we see that the finite size effects in both the Bethe Lattice and the 4D system are qualitatively similar and that the data approach the theoretical curves as the volume becomes large.}

\begin{figure}[t]
\includegraphics[width=1.0\linewidth]{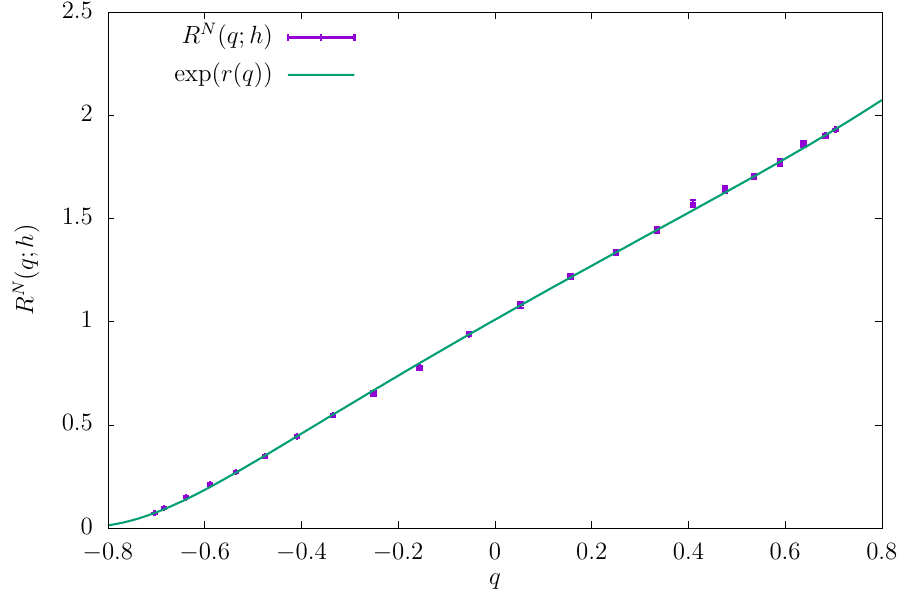}
 \caption{An example of the analytic predictions for the function $R^N(q;h)$ obtained using the function $P(q)$ for the Bethe lattice, with the same parameters as in Fig. 2 of the main article. 
We show the interpolating function $\exp(r(q))$ and the theoretical $R^N(q;h)$ (including the statistical error) for $|q|<q_\mathrm{EA}\approx 0.72$,  and the extrapolations in the region $q_\mathrm{EA}<|q|<0.8$. 
 \label{fig:Extrapolation}}
\end{figure}

We recall that we have plotted the extrapolated predictions also in the region beyond the peaks. A  fit to $\exp(r(q))$ with $r(q)=(a+bq+cq^2+dq^5)/(1+q)$ and the extrapolation in the region $q_\mathrm{EA}<|q|<0.8$ is shown in Fig.~\ref{fig:Extrapolation}. This procedure is somewhat arbitrary, but it affects only the behavior in the tails at $|q|>q_\mathrm{EA}$, i.e. a region that shrinks to zero in the infinite volume limit.
\section*{SI:A detailed comparison of analytic predictions and the numerical data} 
\subsection*{The functions $P$'s and $R$'s}

\begin{figure}[t]
\includegraphics[width=1.0\linewidth]{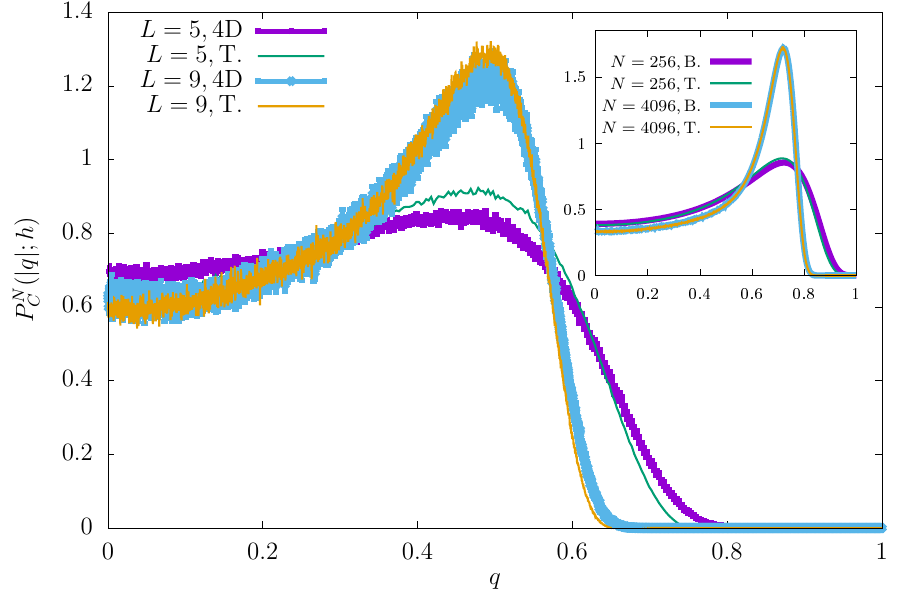}
 \caption{The function $P_C^N(|q|;h)$ versus $q$ at $T=0.7 \; T_c$ for $h=10$ and different values of $L$ in the 4D EA model. Lines correspond to the theoretical prediction from the ultrametric tree (T.) and points with error bars correspond to simulations. Inset: The function $P_C^N(|q|;h)$ against $q^2$ at $T=057 \; T_c$ for $h=4$ and different values of $N$ in the Bethe lattice. 
 \label{fig:PC}}
\end{figure}

\begin{figure}[t]
\centering
\includegraphics[width=1.0\linewidth]{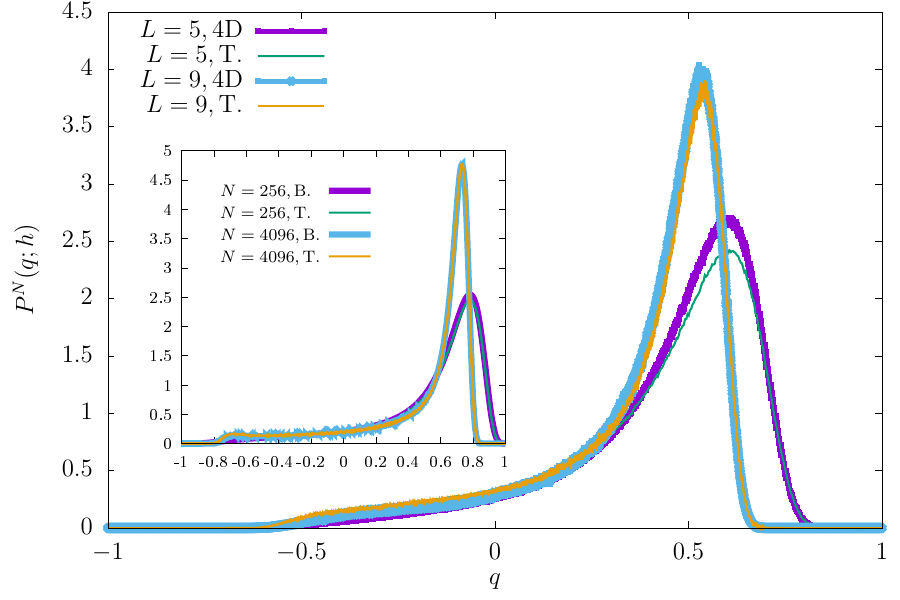}
 \caption{The function $P^N(q;h)$ versus $q$ at $T=0.7\; T_c$ for different values of $L$ in 4D EA model ($h=10$). Inset: $P^N(q;h)$ as a function of $q$ at $T=0.7\; T_c$ for different values of $N$ in Bethe lattice ($h=4$).
 \label{fig:PH}}
\end{figure}

In Figs. \ref{fig:PC} and \ref{fig:PH} we show the functions $P_C^N(|q|;h)$ and $P^N(q;h)$, respectively.

We see that for both functions, for small size, the analytic computations are less accurate but qualitatively correct: this should be expected as far as the analytic computations are exact only in the infinite volume limit. Moreover, the peaks are wider, and the tail of the peaks is quite large: in this situation, our fitting procedure for the function $P(q)$ has a higher degree of approximation. The important message is that the difference between theory and numerical data strongly decreases by increasing the volume.

\begin{figure}[t]
\centering
\includegraphics[width=1.0\linewidth]{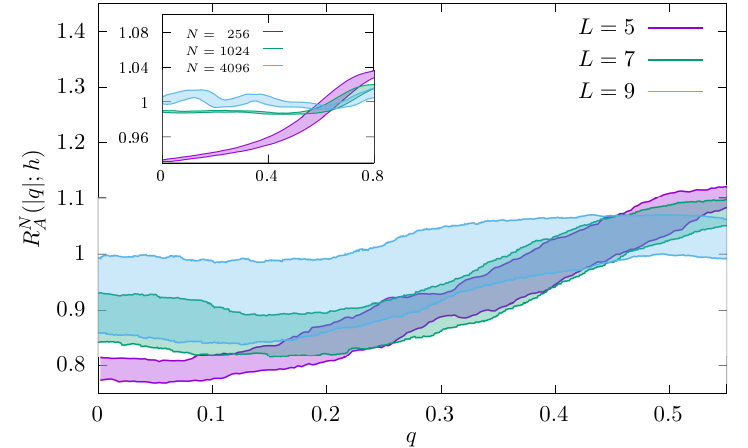}
 \caption{The ratio $R_A^N(|q|;h)$ as a function of $q$ at $T=0.7\; T_c$ for different values of $L$ in 4D EA lattice ($h=10$). Inset: ratio $R_A^N(q;h)$ as a function of $q$ at $T=0.5\; T_c$ for different values of $N$ in Bethe lattice ($h=4$). 
 \label{fig:One}}
\end{figure}

 In order to check if the general prediction that the function $P^N(|q|;h)$ does not depend on $h$, in Fig.\ref{fig:One} we have plotted the ratio:
 \begin{equation}
     R_A^N(|q|;h)=\frac{P^N(|q|;h)}{P^N(|q|;0)}\,.
 \end{equation}
 This ratio should be one in the infinite volume limit: we see that for smaller volumes the ratio is near but not exactly one, but approaches one for larger volumes. A peculiar property of RSB is that $R_A^N(|q|;h)$ is independent of $h$ while $R_C^N(|q|;h)$ is strongly dependent on $h$.

 The reader may notice that we have plotted transparent bands instead of points in Fig.~\ref{fig:One}. We have, for each of the possible values of $|q|$, a value of the observable $R_A^N(|q|;h)$. The problem with plotting the points is that the value of $R_A^N(|q|;h)$ is too noisy and curves for different $L$'s (or different $N$'s for the Bethe case) overlap each other, hindering the visibility. 

 To solve this situation, we have defined a window around each value of $|q|$ and we have computed the mean $\mu_q$ and the standard deviation $\sigma_q$ for the points inside that window. Then, we have plotted the transparent bands of Fig.~\ref{fig:One} with limits $\left[ \mu_q - 2 \sigma_q , \mu_q + 2\sigma_q \right]$.
 
\begin{figure}[t]
\centering
\includegraphics[width=0.8\linewidth]{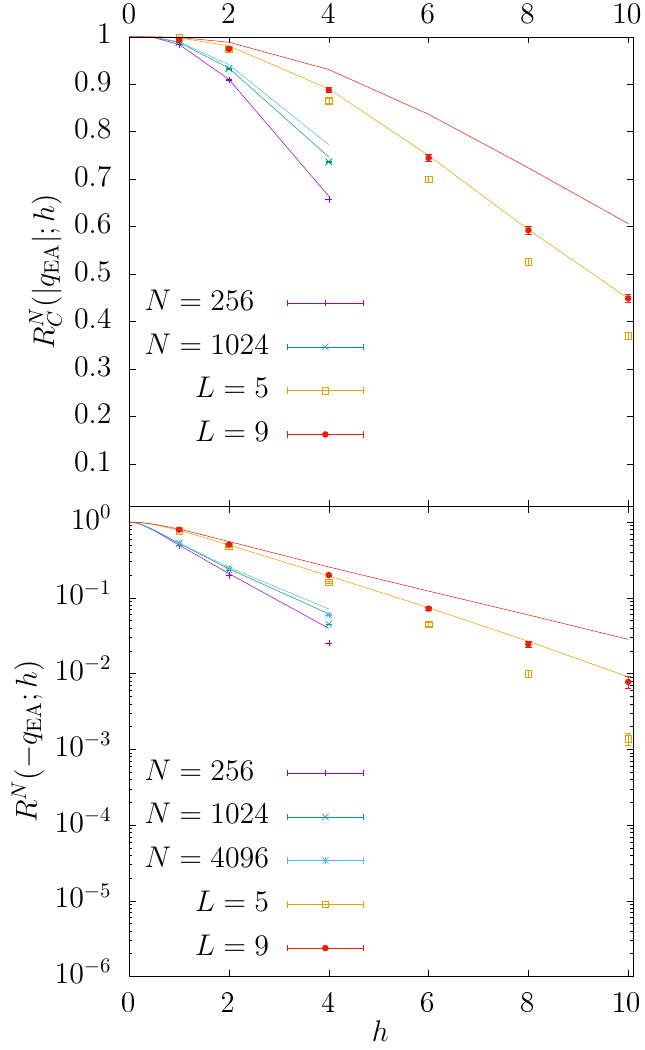}
cp  \caption{Top: The ratio $R_C^N(|q_\mathrm{EA}|;h)$  at $T=0.7T_c$ for different values of $N$ in the Bethe lattice and of $L$ in 4D EA. Lines correspond to the theoretical prediction. Bottom: The ratio $R^N(-q_\mathrm{EA};h)$  at $T=0.7T_c$ for different values of $N$ in the Bethe lattice and of $L$ in the 4D EA. Lines correspond to the theoretical prediction. \label{fig:PEA}}
  
\end{figure}

In Figs. \ref{fig:PC} and \ref{fig:PH} we show the results for a given value of $h$.  To study the dependence on the value of the magnetic field we focus on the values at $|q|=q_\mathrm{EA}$: 
we depict in Fig. \ref{fig:PEA}   the ratios $R^N(-q_\mathrm{EA};h)$ and  $R_C^N(|q_\mathrm{EA}|;h)$ as a function of $h$ \footnote{The values of $R^N(-q_\mathrm{EA};h)$ follow very roughly the behavior $R^N(-q_\mathrm{EA};h)=\exp(-Ah^\delta)$ with $\delta\sim 1.3$ for 4D lattice and $\delta\sim 1.1$ for Bethe. To study how the factor A scales with $N$ we have tried a power law behavior  $A\propto N^b$, obtaining $b=-0.019(19)$ for 4D and $b=-0.047(9)$ for the Bethe lattice.}.  Theory and simulations qualitatively agree also for small systems, and their difference decreases when increasing the volume.  Both quantities should go to zero asymptotically at large $h$. A detailed analysis~\cite{franz:24} tells us that the chaos ratio $R_C^N(|q_\mathrm{EA}|;h)$ goes to zero as $\exp (- A h^2)$. Unfortunately, numerically it is hard to see that behavior. We need to sample the region
$h\gg 1$: for not too large $N$ this requirement conflicts with the 
need to stay at a small total magnetic field $H=h N^{-1/2}$ as required by our simulations (see the large finite-size effects that are present already at the not-very-large fields we have used). Also, the analytic techniques we use (random generators of the trees) are not well suited for large $N$ because it would require the generation of an exponentially large number of leaves.

\subsection*{Large deviations in the negative tails}

To study the large deviations in the negative tails of $P^N (q,h)$ we design the following procedure. Firstly, let us define
\begin{equation}
   I_J(h) = \int_{-q_\mathrm{EA}-\Delta}^{-q_\mathrm{EA}+\Delta} \dd q~P_J^N(q,h)\,.
\end{equation}
Notice that we have an integral $I_J(h)$ for each $J$-sample and magnetic field.
Next, for a given value of $h$, we sort these integrals from the lowest value to the highest one, denoting the newly ordered set of integrals (as a function of the sample number)  as $I_{J^o}(h)$. Finally, we define
\begin{equation}
    R_{<,{J^o}}^N(-q_\mathrm{EA};h) = \frac{I_{J^o}(h)}{I_{J^o}(0)} \,.
    \label{eq:Negative}
\end{equation}
We remark that each $h$ has its own ordered set of couplings $J^o$, and that this order could change from different values of $h$, including $h=0$.

With these  $R_{<,{J^o}}^N(-q_\mathrm{EA};h)$ values we compute the cumulative probability distribution $\mathrm{Prob}( R_{<,{J^o}}^N(-q_\mathrm{EA};h)<R)$.

We notice the agreement is excellent for not too small $R$ with the exclusion of the highest fields. However, for this observable for $h=10$ the numerical data show a very strong dependence on the size of the system.

If we look at small values of $R$ we find that the numerical cumulative is higher than the analytic predictions: in other words, the tail at smaller values of $R$ is higher than the theoretical predictions, i.e. we miss samples in the region of $0<R<.1$ and we have an excess of samples with $R=0$. These results should not be surprising: we have run the simulations up to 30 times the thermalization time in our thermalization criteria. Consequently, regions of phase space with a probability lower than 1/30 may be missed. It is quite possible that the discrepancy between theory and numerical simulations at not too large $h$ would strongly decrease with much longer simulations.

\begin{figure}[t]
\centering
\includegraphics[width=1.0\linewidth]{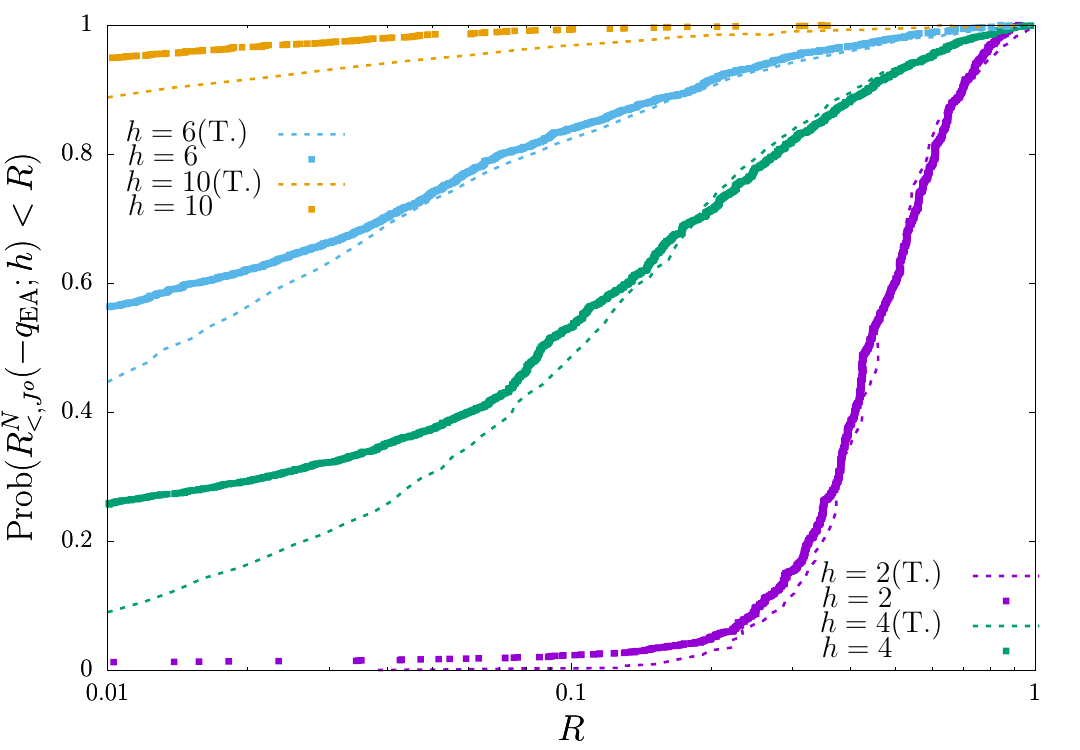}
\caption{Theory and 4D EA data for $\mathrm{Prob}( R_{<,{J^o}}^N(-q_\mathrm{EA};h)<R)$ as a function of $R$, 
for $h=2,4,6$ and 10 and $L=9$. Dashed lines correspond with the theoretical prediction.}
 \label{fig:LARGE}
\end{figure}

In Fig. \ref{fig:LARGE} we confront the numerical data from the 4D EA model with the theoretical result for this cumulative probability. We show data for $h=2, 4, 6$, and 10 and for the largest simulated lattice $L=9$. We find good agreement for small and intermediate values of 
$h$ which deteriorates for very large values of $h$.

\end{document}